\documentclass[10pt,prl,aps,reprint,graphicx,showpacs]{revtex4-1}
\usepackage{graphicx}
\usepackage{amssymb}
\usepackage{subfigure}
\begin{document}

\title{Resistance minimum and electrical conduction mechanism in polycrystalline CoFeB thin films}

\author{G. Venkat Swamy$^{1,2}$, P. K. Rout$^{2,\ast}$, Manju Singh$^2$, and R. K. Rakshit$^{1,2}$}

\affiliation{$^1$Academy of Scientific and Innovative Research, CSIR-National Physical Laboratory Campus, Dr. K. S. Krishnan Road, New Delhi - 110012, India\\
$^2$CSIR-National Physical Laboratory, Dr. K. S. Krishnan Road, New Delhi - 110012, India}
\email{pkrout@mail.nplindia.org}

%\date{\today}

\begin{abstract}
The temperature dependent resistance $R$($T$) of polycrystalline ferromagnetic CoFeB thin films of varying thickness are analyzed considering various electrical scattering processes. We observe a resistance minimum in $R$($T$) curves below $\simeq$ 29 K, which can be explained as an effect of intergranular Coulomb interaction in a granular system. The structural and Coulomb interaction related scattering processes contribute more as the film thickness decreases implying the role of disorder and granularity. Although the magnetic contribution to the resistance is the weakest compared to these two, it is the only thickness independent process. On the contrary, the negative coefficient of resistance can be explained by electron interaction effect in disordered amorphous films.\\
\\
Keywords: Ferromagnetic alloy, thin film, spintronics material
\end{abstract}


% start main body here................
%\clearpage
%\baselineskip=1.5cm
\maketitle

The ferromagnetic CoFeB has been extensively employed in various spintronics devices like magnetic tunnel junctions \cite{S. Ikeda1,Wang,S. Ikeda2}, spin valves \cite{C. Y. You}, and spin-torque devices \cite{A. A. Tulapurkar,Luqiao Liu} over the last decade. The electrical conduction in these devices utilising the spin degree of freedom depends on the composition, spin polarization, and crystallinity of the CoFeB layers. Therefore, a basic understanding of the electrical and magneto-transport properties of CoFeB thin films prepared under different conditions is essential for further improvement of any device based on this ferromagnetic alloy. Still there are only few reports on the electrical transport properties of CoFeB thin films including composites and nanotubes \cite{Fujimori,P. Johnsson,S. U. Jen,Deng,S. X. Huang,G. V. swamy,Ruffer}. The electrical resistivity {$\rho$($\emph{T}$)} of a crystalline film is significantly lower as compared to that of an amorphous film \cite{S. X. Huang,G. V. swamy}. While the crystalline films show a metallic behaviour, negative temperature coefficient of resistance has been reported for the amorphous or composite films \cite{P. Johnsson,G. V. swamy}. Thus, the crystalline quality plays an important role in the process of electrical conduction of these films. Moreover, the thickness of these films also controls their electrical properties as the size effects become dominant at lower thicknesses. For example, Jen \textit{et al.} have reported increase of resistivity in the amorphous CoFeB film on lowering of the thickness \cite{S. U. Jen}. However, a thorough analysis of electrical transport in crystalline CoFeB films is lacking.

An interesting feature of CoFeB systems is the presence of resistance upturn at lower temperatures \cite{Fujimori,G. V. swamy,Ruffer,Kettler}. However, there is no clear consensus on its origin due to the differences in experimental results and theoretical interpretations. Fujimori \textit{et al.} have reported a ${e^{ - \sqrt T }}$ dependent resistivity in the upturn region \cite{Fujimori}, which is expected for a variable range hopping conductivity in Coulomb gap \cite{Efros}. A logarithmic $T$ dependent resistivity has been observed in crystalline CoFeB film \cite{G. V. swamy} and nanotube \cite{Ruffer} as well as amorphous ribbons \cite{Kettler}, which can originate due to one or more of these three effects; Kondo scattering \cite{Kondo} or Coulomb effect in granular materials \cite{Efetov} or tunneling model \cite{Cochrane}. The resistance minimum in variety of amorphous ferromagnetic alloys similar to CoFeB has been attributed to Kondo type $lnT$ resistance \cite{Hasegawa1,Hasegawa2,Barquin,Shen} and, in some alloys, an additional minimum due to $T^{1/2}$ resistance has been observed \cite{Olivier}. Moreover, in a strong ferromagnetic material like CoFeB, the role of magnon in electronic transport is quite important apart from other scattering mechanisms involving electron, phonon, lattice potentials due to structure etc. For example, a finite magnetic contribution to the resistivity due to electron-magnon scattering along with the structural contribution has been observed in amorphous ferromagnetic alloys like FeBC and FeBGe \cite{Kaul2,Kaul1}. All these mechanisms mostly result in various power law temperature dependence of resistivity. In order to shed some light on these aspects, we have analyzed the temperature dependent electrical transport in polycrystalline Co$_{40}$Fe$_{40}$B$_{20}$ thin films of varying thickness in detail. We have mainly focussed on the origin of resistance minimum observed below $\simeq$ 29 K in the light of various scattering mechanisms involved in the electrical conduction.

The thin films of Co$_{40}$Fe$_{40}$B$_{20}$ with nominal thicknesses ranging from 5 to 40 nm have been deposited on SiO$_2$(300 nm)/Si(100) substrates using KrF excimer ($\lambda$ $\approx$ 248 nm) based pulsed laser deposition technique. The growth of these films has been performed at room temperature under argon pressure of 2 $\times$ 10$^{-3}$ mbar with a typical growth rate of 0.04 nm/s. To enhance the crystallinity, the films have been subsequently annealed at 400$^\circ$C for 1 hour under high vacuum. The details of film growth have been reported previously \cite{G. V. swamy}. Figure 1 shows the grazing angle X-ray diffraction scans for various CoFeB films. We have determined the actual thickness of the films from the fits to low angle scans or X-ray reflectivity (XRR) as shown in Fig. 1(b). The extracted thickness ($t$) of the films are 6, 9, 20, and 36 nm. The high angle scan profiles show the polycrystalline nature of the films [See Fig. 1(a)]. As the film thickness increases, the intensity of (110) CoFe peak increases and it becomes much sharper as shown in Fig. 1(c), indicating better crystalline quality and the presence of larger grains in thicker films. While the fits to XRR data provide an estimation of interface roughness values as $\leq$ 1 nm, the surface roughness obtained from atomic force microscopy (AFM) scans is $\approx$ 1.5 nm. Moreover, the AFM images reveal the granular nature of these films as shown in Fig. 2(a). The grain size decreases with decreasing film thickness. Similar trend can be observed from the average grain size ($L$) determined by the Scherrer formula [See Fig. 2(b)]. We have previously reported the presence of grains with similar sizes as observed by transmission electron microscopy imaging of CoFeB films\cite{G. V. swamy}. This variation of grain size with film thickness has significant effects on magnetic and transport properties as we will discuss now.

\begin{figure}
\begin{center}
%\vskip 2cm
%\abovecaptionskip -1cm
\includegraphics[width=8cm]{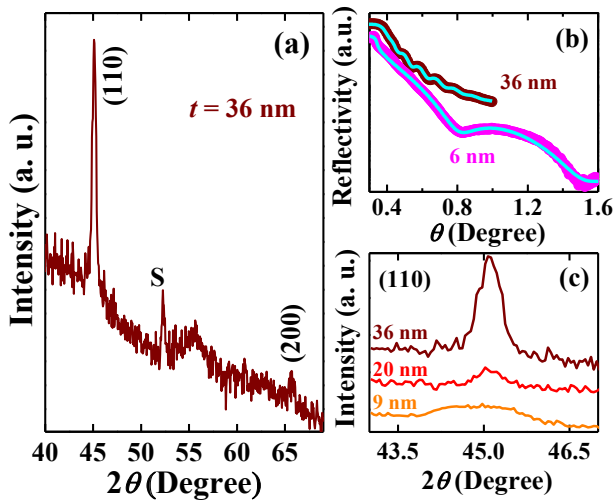}%,
\end{center}
%\vskip -4cm
\caption{\label{Fig1} (Color online) (a) X-ray diffraction pattern of 36 nm thick CoFeB film. The substrate peak is denoted by ``S". (b) X-ray reflectivity scans for two films with $t=$ 6 and 36 nm along with the fits (solid lines). (c) X-ray diffraction pattern around (110) CoFe peak for 9, 20, and 36 nm thick films.}
\end{figure}

\begin{figure}
\begin{center}
%\vskip 2cm
%\abovecaptionskip -1cm
\includegraphics[width=8cm]{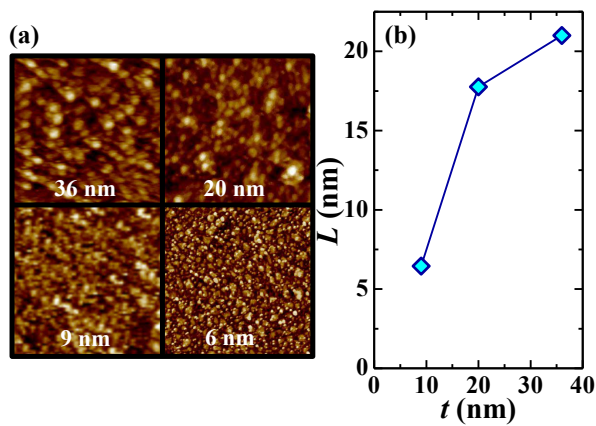}%,
\end{center}
%\vskip -4cm
\caption{\label{Fig2} (Color online) (a) The AFM images (1 $\times$ 1 $\mu$m$^2$ area) for all films. (b) The grain size ($L$) as a function of thickness.}
\end{figure}

\begin{figure}
\begin{center}
%\vskip 2cm
%\abovecaptionskip -1cm
\includegraphics[width=8.5cm]{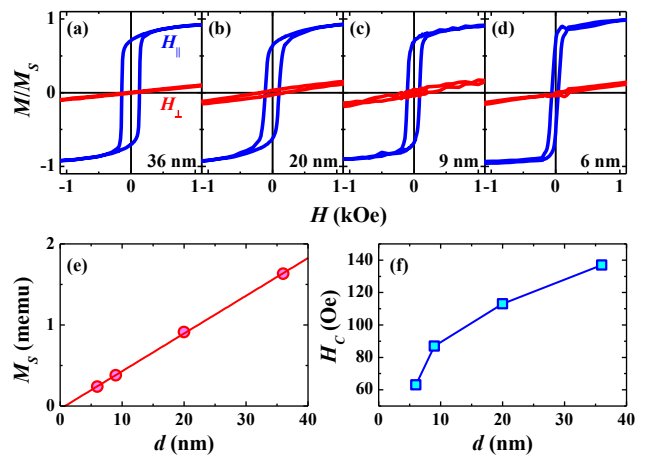}%,
\end{center}
%\vskip -4cm
\caption{\label{Fig3} (Color online) (a-d) The in-plane and out-of-plane magnetic hysteresis loops for the films with $t$ = 6, 9, 20, and 36 nm measure at $T$ = 300 K. The thickness dependence of saturation magnetic moment ($M_S$) and coercivity ($H_C$) are shown in the panel (e) and (f), respectively.}
\end{figure}

Figure 3(a-d) display the room temperature magnetic hysteresis loops of all films measured for both in-plane and out-of-plane fields. Clearly, we observe low saturation fields and high squareness ratio for in-plane loops. While the in-plane squareness is greater than 0.7, the out-of-plane squareness is less than 0.05. These observations indicate that the magnetic easy axis is in the film plane. The saturation magnetization of CoFeB can be determined from the linear fit to the thickness dependent saturation magnet moment ($M_S$). The extracted value of saturation magnetization comes out to be 1580 emu/cc. Furthermore, we observe a magnetic dead layer of 0.8 nm, which is close to the values of dead layer observed in other CoFeB interfaces \cite{S. Ikeda2}. Figure 3(f) presents the coercivity ($H_C$) of easy axis loop as a function of $d$, which clearly shows the enhancement of $H_C$ with increasing $d$. Similar behaviour observed in various ferromagnetic alloys has been attributed to the grain size effect \cite{W.S. Sun, Yuan-Tsung Chen, Dongsheng Yao}. With decreasing grain size, the effective anisotropy constant reduces, which leads to increase in effective exchange range\cite{Dongsheng Yao}. We have also observed such reduction in anisotropy with decreasing thickness (or decreasing grain size) \cite{G. Venkat Swamy2}. Under such scenario, the exchange-coupling between the grains is enhanced and, thus, the $H_C$ reduces with decreasing thickness .

Now, we present the electrical transport study of our films [See Fig. 4(a)]. The temperature dependent four-probe sheet resistance $R_{\square}$($T$) shows the metallic behaviour for all CoFeB films. The resistance decreases with lowering of the temperature up to a certain temperature, known as resistance minimum temperature ($T_{min}$), and then it starts increasing down to 5 K. The $T_{min}$ for the 36 nm thick film is 26 K and it gradually increases with decreasing thickness [See Fig. 4(b)]. The thickness dependence of sheet conductance $G_{\square} = 1/R_{\square}$ at $T =$ 273 K shows a linear thickness dependence as shown in Fig. 4(c). The slope of the linear fit yields a resistivity of (33 $\pm$ 2) $\mu\Omega$ cm, which is close to previously reported values for several micron thick amorphus CoFeB ribbons \cite{Kettler}. However, one can observe a finite intercept of the linear fit on the temperature axis, which suggests the presence of an electrically less conducting dead layer. The existence of such dead layers has been reported in thin films of manganite \cite{Sun} and Heusler alloy \cite{Rout}. This dead layer can form on the surface and at film-substrate interface due to surface oxidation, roughness effect, and boron segregation at the interface \cite{S. U. Jen,G. V. swamy}. Alternatively, the presence of larger number of grain boundaries in thinner films can also result in such behavior. The thickness of the dead layer is given by the intercept on the thickness axis, which comes out as $t_d \simeq$ 2.5 nm.

\begin{figure}
\begin{center}
%\vskip 2cm
%\abovecaptionskip -10cm
\includegraphics[width=8.5cm]{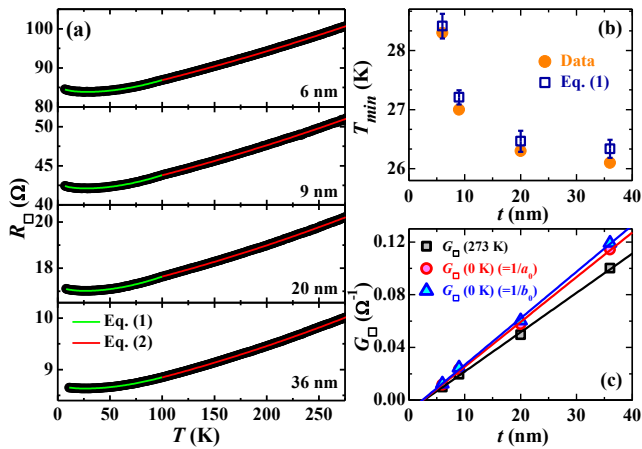}%,
\end{center}
%\vskip -3.5cm
\caption{\label{Fig4} (Color online) (a) Temperature dependent sheet resistance $R_{\square}$($T$) plots for CoFeB ($t =$ 6, 9, 20, and 36 nm) films. The solid lines are the fits to $R_{\square}$($T$) according to Eq. (1) and (2) with the quality of fitting given by $r^2 >$ 0.999. (b) Thickness dependence of $T_{min}$, which is the temperature at which d$R_{\square}$($T$)/d$T =$ 0 and $T_{min} = ( - \beta /2{a_2})^{1/2}$ using the fitting parameters in Eq. (1). (c) Thickness dependence of $G_{\square}$ at $T =$ 273 K along with $G_{\square}$(0 K) obtained from $a_0$ and $b_0$. The straight lines are the linear fits to the data.}
\end{figure}

Coming back to $R_{\square}$($T$) data, we have fitted these using the following expressions. For low temperature (5 K $\leq T \leq$ 100 K) regime, the resistance is expressed as:
\begin{eqnarray}
{R_{\square}(T)} ={a_0} + {a_2} T ^ 2 + {\beta} ln T,
\end{eqnarray}
and, for high temperature (100 K $\leq T \leq$ 273 K) regime,
\begin{eqnarray}
{R_{\square}(T)} ={b_0} + {b_1} T + {b_2} T ^ {2},
\end{eqnarray}
where $a_0$, $a_2$, $\beta$, $b_0$, $b_1$, and $b_2$ are the constants. The temperature independent terms ($a_0$ and $b_0$) represent the residual sheet resistance $R_{\square}$(0 K). Using these values, the $t_d$ again comes out as $\simeq$ 2.5 nm [See Fig. 4(c)], which implies that the dead layer remains fixed in whole temperature range. The $T^2$ term in above expressions can arise due to the effects like electron-magnon scattering and electron-lattice scattering. The first scattering process is due to the coherent scattering of electrons by long-wavelength magnons in a ferromagnet, which introduces a magnetic contribution ($\rho_{mag}/t$) to $R_{\square}$. On the other hand, the incoherent scattering can introduce a $T^{3/2}$ dependent term in an amorphous ferromagnet while such process is absent in a crystalline ferromagnet \cite{Kaul2}. Therefore, we do not observe a $T^{3/2}$ dependency in our fits. The latter process involves the scattering of conduction electrons from the lattice potential of a transition-metal system, which results in a structural contribution ($\rho_{str}/t$) to $R_{\square}$ \cite{Cote1,Cote2}. While $\rho_{mag}$ varies as $T^2$ for all temperatures, $\rho_{str}$ shows a transition from $T^2$ to linear $T$ dependence as the temperature increases. The last term in Eq. (1) can arise due to two effects. Firstly, the logarithm dependence of resistance is a characteristics feature of Kondo scattering observed in dilute magnetic metals, quantum dots, and heavy electron systems \cite{Kondo}. The Kondo behavior can be explained as the interaction of conduction electrons with localized spins of magnetic impurities. Such effect can also be observed in ferromagnetic materials, where the effective field distribution for a magnetic spin has a long tail extended below zero field and, thus, some of the spins participate in spin flipping scattering process \cite{Grest}. One should note that this effective field model is only valid for amorphous systems while, in crytalline ferromagnet, distinct field lines are observed in place of a field distribution \cite{Chien}. Also, the absence of resistance saturation (or tendency for saturation) in many reports does not provide a solid evidence of Kondo effect \cite{Kettler,Barquin,Shen}. Thus, for our metallic crystalline ferromagnetic CoFeB thin films, the presence of Kondo effect well below its Curie temperature (\emph{T}$_c$) is highly unlikely. We will provide further support to our conjecture later on. The second possible explanation is the Coulomb interaction in a granular metal, where a logarithmic temperature dependent conductivity can be observed in metallic regime with dimensionless tunneling conductance $g \gg 1$ \cite{Efetov,Beloborodov}. To verify if this scenario is valid in our case, we have determined $g$. In the temperature range of 5 K $\leq T \leq$ 100 K, the maximum relative change in resistance, i.e., [$R_{\square}$(100 K) - $R_{\square}$(${T_{\min }}$)]/$R_{\square}$(${T_{\min }}$) is $<$ 0.04. For such small variation, the resistivity can be approximated to: $\rho \sim (\alpha /{\sigma _0})\ln (g E_c / T)$, where $\alpha  = {(2\pi gd)^{- 1}}$, ${\sigma _0} (= {e^2}gL^{2-d}/\hbar)$ is the conductivity of the granular metal without the Coulomb interaction, and $E_c$ is the charging energy of an isolated metal grain. Here, $d$ is the dimensionality of the granular array, which corresponds to the half of the number of neighboring grains to a single grain rather than to real dimensionality. Comparing this expression with the logarithmic term in Eq. (1), we have $\beta  =  - (\hbar /2\pi d{e^2}){(L/t)^{d - 2}}{g^{ - 2}}$. Figure 5 shows the estimated values of $g$ for $d =$ 2 and 3 using the fitting parameter $\beta$. Clearly, $g \gg$ 1 for all the films, which indicates that the conduction is not a tunneling process through an insulator; rather a charge transfer process between the grains via metallic grain boundaries as observed in many metallic alloys \cite{Xiao1,Xiao2}. Moreover, the $-\beta t$ increases with decreasing thickness. This can be explained in terms of the presence of more disorders and grains in thinner films, which results in reduction of $g$ and, thereby, enhancement of $-\beta t$. All these observations indicate that the resistance minimum observed here can be due to the granular nature of the films and the enhancement of $T_{min}$ in thinner films is due to increasing granularity and disorder. We can also determine the $T_{min}$ using Eq. (1) as $T_{min} = ( - \beta /2{a_2})^{1/2}$. These values match quite well with experimental values as shown in Fig. 4(b), which is an indicator of good quality of these fits.

\begin{figure}
\begin{center}
%\vskip 2cm
%\abovecaptionskip -10cm
\includegraphics[width=8.5cm]{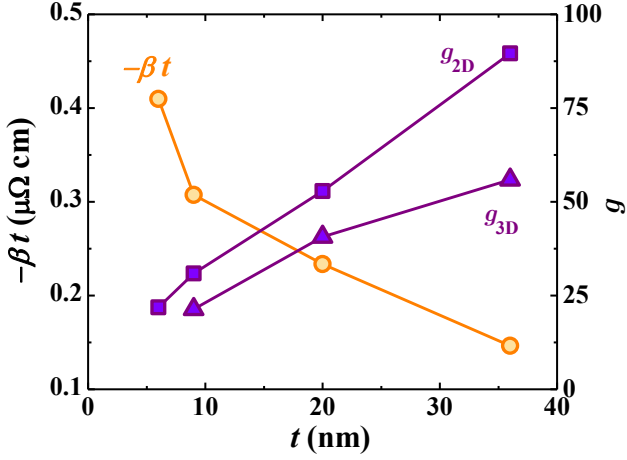}%,
\end{center}
%\vskip -3.5cm
\caption{\label{Fig5} (Color online) Thickness dependence of -$\beta t$ as well as $g$ for two and three dimensional granular array.}
\end{figure}

\begin{figure}
\begin{center}
%\vskip 2cm
%\abovecaptionskip -10cm
\includegraphics[width=8.5cm]{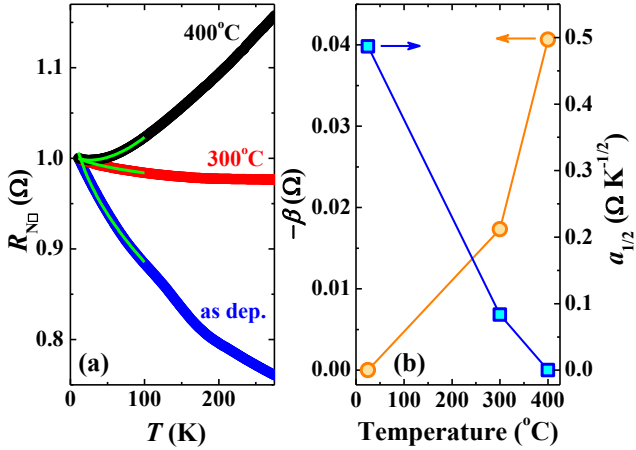}%,
\end{center}
%\vskip -3.5cm
\caption{\label{Fig6} (Color online) (a) The normalised sheet resistance $R_{N\square}$($T$) $=$ $R_{\square}$($T$)/$R_{\square}$(10 K) of as deposited as well as annealed 40 nm films. The curves are fitted to the expression: ${R_{\square}(T)} ={a_0} + {a_{1/2}} T ^ {1/2} + {a_2} T ^ 2 + {\beta} ln T$. The fitting parameters $a_{1/2}$ and $-\beta$ are plotted in the panel (b).}
\end{figure}

We have analysed the normalized sheet resistance $R_{N\square}$($T$) of as deposited and annealed 36 nm thick films to get further insight in to the observed resistance minimum [See Fig. 6(a)]. In case of as-deposited film, we observe a negative temperature coefficient of resistance (TCR) with a residual resistance ratio RRR $= R_{N\square}$(273 K) $\simeq$ 0.76. Upon annealing the film at 300$^\circ$C, the RRR increases to $\simeq$0.98 while negative TCR still persists for whole temperature range. Further annealing at 400$^\circ$C increases the metallically of the film with RRR $\simeq$ 1.15 and negative TCR regime present below 26 K. Such enhanced metallicity with thermal annealing is directly related to enhancement of crystallinity in these films \cite{G. V. swamy}. While the 400$^\circ$C annealed film is granular, the TEM image of as deposited film show a feature-less amorphous matrix with scarcely dispersed crystalline regions. So we do not expect any $ln T$ contribution coming due to inter-granular Coulomb interactions in as deposited film. On the other hand, a Kondo-type $ln T$ dependent resistance is expected for this amorphous film, where the effective field model is valid and this effect should reduce with annealing as the film becomes gradually crystalline. In contrast, the $R_{N\square}$($T$) of this film does not fit to Eq. (1), which again points against Kondo effect in this system. We have tried to fit $R_{N\square}$($T$) by Eq. (1) with an additional $T^{1/2}$ term, which represents the electron-electron ($e$-$e$) interactions in a three dimensional disordered system \cite{Lee}. With this new expression, negative TCR in the as deposited film can be explained by only $T^{1/2}$ term while the $lnT$ term vanishes [See Fig. 6(b)]. In case of 300$^\circ$C annealed film, the $R_{N\square}$($T$) has contribution from both $ln T$ and $T^{1/2}$ terms. The relative weight of these terms, i.e. $W({a_{1/2}},\beta ) = \int {{a_{1/2}}{T^{1/2}}} dT/\int {\beta \ln T} dT$ within the limits 10 K $\leq T \leq$ 100 K, comes out as $\simeq$9.0, which indicates the $e$-$e$ interaction is still highly dominant process for 300$^\circ$C annealed film. Olivier \textit{et al.} have reported the presence of both $lnT$ and $T^{1/2}$ dependent resistance in FeCrB metallic glasses, where the $T^{1/2}$ term is responsible for resistance minimum in some cases \cite{Olivier}. Thus, the presence of negative TCR in all three samples (albeit in different temperature regimes) has two different origins; i.e. $e$-$e$ interactions in a disordered amorphous film and Coulomb effect in granular polycrystalline film.

In order to find out the significance of quadratic temperature dependent resistance, we have converted the coefficients ($a_2$ and $b_2$) in the form of resistivities, i.e., ${a'_2} = a_2 t$ and ${b'_2} = b_2 t$. The ${b'_2} T^2$ term represents the $\rho_{mag}$ while the ${a'_2} T^2$ term is a combination of $\rho_{mag}$ and $\rho_{str}$. Figure 7(a) shows that ${b'_2}$ (or $\rho_{mag}$) remains almost constant irrespective of $t$. Such behaviour implies that all films are magnetically homogeneous, which can also be confirmed from magnetic hysteresis loop measurements shown in Fig. 3. Moreover, the disorders present in thinner films only introduce a minor correction to the magnetic $T^2$ term and, thus, do not significantly alter $\rho_{mag}$ \cite{Kaul2}. The ${b'_2}$ values are quite close to the values obtained for other ferromagnetic metals (2.2 - 3.2 $\times$ 10$^{-11}$ $\Omega$ cm K$^{-2}$ for Fe and Co) and alloys (0.98 $\times$ 10$^{-11}$ $\Omega$ cm K$^{-2}$ for Fe$_{80}$B$_{20}$) \cite{Kaul2,Raquet,Goodings}. The $\rho_{mag}$ is proportional to ${({J_{sd}}/D)^2}$, where $J_{sd}$ is the $s$-$d$ exchange integral and $D$ is the spin-wave stiffness constant \cite{Kaul2}. We can estimate the values of $D$ from temperature dependent magnetization $M(T)$ data as shown in the inset of Fig. 7(a). The $M(T)$ in the low temperature regime is given by Bloch law as follows: $M(T) = M(0)(1 - B{T^{3/2}})$, where $B = [\zeta (3/2)g_e{\mu _B}/M(0)]{({k_B}/4\pi D)^{3/2}}$ \cite{Fernandez-Baca}. Here, $\zeta (3/2) =$ 2.612 is the Riemann $\zeta$ function, $g_e =$ 2 is the gyromagnetic ratio, and $\mu _B$ is the Bohr magneton. Using these expressions, the $D$ comes out as 88 and 185 meV {\AA}$^2$ for the 9 and 36 nm thick films, respectively, which are comparable to the reported values for various FeB based alloys \cite{Fernandez-Baca}. Since the $\rho_{mag}$ [and thereby ${({J_{sd}}/D)^2}$] is thickness independent, $\mid J_{sd} \mid$ ($t =$ 9 nm) $<$ $\mid J_{sd} \mid$ ($t =$ 36 nm). Here, we want to point out that the values of $J_{sd}$ are in negative. In a Kondo picture, $\beta t \propto J_{sd}^3$, which implies that $-\beta t$ ($t =$ 9 nm) $<$ $-\beta t$ ($t =$ 36 nm). But an opposite trend observed in our case [See Fig. 5] discards Kondo effect as the explanation of the resistance minimum. On the other hand, the low temperature $T^2$ coefficient ${a'_2}$ is much larger than ${b'_2}$ and increases with reducing $t$, which suggests an additional contribution apart from thickness independent ${b'_2}$. This extra contribution corresponds to $\rho_{str}$ and it gets enhanced in thinner films due to the presence of more structural disorders and grain boundaries. Comparing these two effects, we observe a substantial magnetic contribution although $\rho_{str}$ is at least 3.5 times more dominating in comparison to $\rho_{mag}$.

\begin{figure}
\begin{center}
%\vskip 2cm
%\abovecaptionskip -10cm
\includegraphics[width=8.5cm]{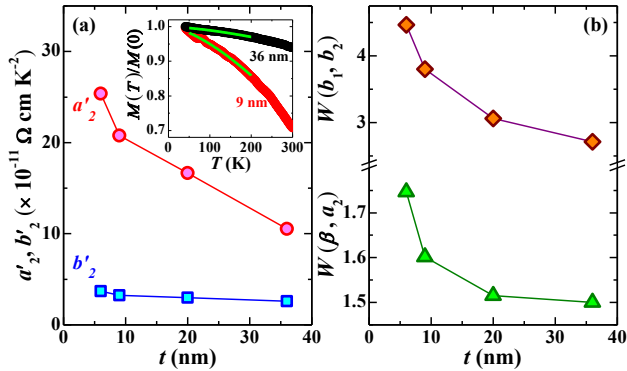}%,
\end{center}
%\vskip -3.5cm
\caption{\label{Fig7} (Color online) Thickness dependence of (a) $a'_2$, $b'_2$, and (b) $W$($\beta$,$a_2$), $W$($b_1$,$b_2$). The inset of the panel (a) shows the $M$($T$)/$M$(0) curves for 6 and 36 nm thick films with the fits given by the relation: $M(T) = M(0)(1 - B{T^{3/2}})$.}
\end{figure}

Apart from the $T^2$ term, the $R_{\square}$($T$) expressions involve two other important terms; i.e. $\beta$ln$T$ (in low $T$ regime) and $b_1 T$ (in high $T$ regime). We have defined the relative weight of these terms with respect to corresponding $T^2$ terms as follows: $W(\beta ,{a_2}) = \int {\beta \ln T} dT/\int {{a_2}{T^2}} dT$ and $W({b_1},{b_2}) = \int {{b_1}T} dT/\int {{b_2}{T^2}} dT$ with the integrations performed using the limits 5 K $\leq T \leq$ 100 K and 100 K $\leq T \leq$ 273 K, respectively. Figure 7(b) shows these relative weights as a function of $t$. Clearly, the Coulomb effect plays a dominating role as compared to the $T^2$ term at lower temperatures as $W(\beta ,{a_2}) > 1$. Moreover, the increase of $W(\beta ,{a_2})$ with decreasing thickness suggests that this effect becomes rather important as the grain size reduces and the grain boundary contribution increases. In high temperature regime, $\rho_{str} \sim T$ and the $W({b_1},{b_2})$ basically represents the relative weight of $\rho_{str}$ with respect to $\rho_{mag}$. Thus, the dominating effect of linear $T$ term and its enhancement with decreasing $t$ are well expected as explained before. While the electron-phonon ($e$-$p$) scattering also shows a linear $T$ dependency above Bloch-Gruneisen temperature ($T_{BG}$), it should reduce with increasing grain boundary and, thus, with decreasing thickness \cite{Boekelheide}. This behaviour is opposite to the trend seen in Fig. 7(b). Moreover, the $e$-$p$ scattering in low $T$ regime ($T << T_{BG}$) results in a $T^5$ dependent resistance term. We have not observed any improvement to the fits by additional introduction of this term in Eq. (1). Therefore, one can safely assume that the electron-phonon scattering does not play an important role in the electronic conduction.

In conclusion, we have investigated pulsed laser deposited polycrystalline CoFeB films of varying thickness. The analysis of transport measurements of CoFeB thin films demonstrates the effect of granularity and disorder on the structural, magnetic, and Coulomb interaction related scattering processes. The resistance minimum is related to the grains (and the grain boundaries) present in the film. While the magnetic contribution to resistance remains independent of the film thickness, the structural contribution and inter-granular Coulomb effect increases with decreasing thickness. In amorphous films, the electrical conduction is mainly dominated by the electron interaction effects in disordered systems. Our comprehensive study of electronic transport in CoFeB films shows that the electrical conduction in thinner films will be affected by granularity (or Coulomb charging effects). Therefore, better performance of spintronics devices can be achieved either by reducing device size to the order of grain size or by developing a better fabrication technique for thin films with large grains.

We thank R. C. Budhani for his valuable comments, K. K. Maurya for XRD measurements, and V. Toutam for AFM measurements. We acknowledge Council of Scientific and Industrial Research (CSIR) $\&$ Department of Science and Technology (DST) for financial support.

%\newpage

\newpage

\end{document}